\documentclass[a4paper,11pt]{article}
\usepackage{jinstpub} 
\usepackage{lineno}
\usepackage{multirow}

\newcommand{\oli}{\overline}

\newcommand{\bm}{\boldmath}

\newcommand{\sig}{\ensuremath{\sigma}}

\newcommand{\pbar}{\ensuremath{\mathrm{\oli p}}} 

\newcommand{\Hbar}{\ensuremath{\mathrm{\oli H}}}
\newcommand{\Hbarplus}{\ensuremath{\mathrm{\oli H}^+}}
\newcommand{\Hminus}{\ensuremath{\mathrm{H}^-}}

\newcommand{\MCPu}{\ensuremath{\mathrm{MCP_{\,up}}}}
\newcommand{\MCPd}{\ensuremath{\mathrm{MCP_{\,\oli H}}}}

\newcommand{\nclu}{\ensuremath{n_{clu}}}
\newcommand{\Sclu}{\ensuremath{\Sigma_{clu}}}
\newcommand{\Npbar}{\ensuremath{\,\mathrm{N_{\pbar}}}}
\newcommand{\Ntr}{\ensuremath{\,\mathrm{N_{tr}}}}
\newcommand{\Dist}{\ensuremath{\,\mathrm{D}}}
\newcommand{\Ntrppb}{\ensuremath{\,\mathrm{<\!N_{tr/\pbar}\!>}}}
\newcommand{\Asens}{\ensuremath{\,\mathrm{A_{sens}}}}

\newcommand{\geant}{\emph{GEANT4}}

\newcommand{\piz}{\ensuremath{\pi^{0}}}
\newcommand{\pip}{\ensuremath{\pi^{+}}}
\newcommand{\pipm}{\ensuremath{\pi^{\pm}}}

\newcommand{\pim}{\ensuremath{\pi^{-}}}

\newcommand{\cm}{\ensuremath{\,\mathrm{cm}}}
\newcommand{\mum}{\ensuremath{\,\mu\mathrm{m}}}

\newcommand{\mm}{\ensuremath{\,\mathrm{mm}}}
\newcommand{\mmsq}{\ensuremath{\,\mathrm{mm^{2}}}}

\newcommand{\mus}{\ensuremath{\,\mu\mathrm{s}}}

\newcommand{\en}{\ensuremath{\,\mathrm{e^-}}}
\newcommand{\enrms}{\ensuremath{\,\mathrm{e^-_{rms}}}}

\title{\bm Monitoring antiproton numbers with a CMOS 
detector in a dense-track environment}

\date{\scriptsize{Ver.1, \today }}

\collaboration{The GBAR Collaboration}

\author{
C.~Regenfus$^{1,*}$,    
P.~Adrich$^{2}$,        
I.~Belosevic$^{3}$,     
F.~Benkel,$^{1}$,
M.~Chung$^{4}$,         
P.~Cladé$^{5}$,
P.~Comini$^{3}$,        
P.~Crivelli$^{1}$,      
P.~Debu$^{3}$,          
A.~Douillet$^{5,6}$,     
S.~Geffroy$^{7}$,       
S.~Guellati-Khelifa$^{5,8}$,     
P.~Guichard$^{9}$,      
P.-A.~Hervieux$^{9}$,   
L.~Hilico$^{5,6}$,      
P.~Indelicato$^{5}$,    
S.~Jonsell$^{10}$,      
J.-P.~Karr$^{5,6}$,      
B.~Kim$^{11}$,          
S.~Kim$^{12}$,          
E.-S.~Kim$^{13}$,       
N.~Kuroda$^{14}$,       
B.~Lee$^{12}$,          
L.~Liszkay$^{3}$,       
D.~Lunney$^{7}$,        
G.~Manfredi$^{9}$,      
B.~Mansoulié$^{3}$,     
V.~Martimort$^{5}$,     
M.~Matusiak$^{2}$,      
V.~Nesvizhevsky$^{15}$, 
F.~Nez$^{5}$,           
K.~Park$^{12}$,         
N.~Paul$^{5}$,          
E.~Perez$^{16}$,        
P.~Pérez$^{3,17}$,         
C.~Roumegou$^{7}$,      
J.-Y.~Roussé$^{3}$,     
F.~Schmidt-Kaler$^{18}$,
K.~Szymczyk$^{2}$,      
T.~A.~Tanaka$^{14,19}$, 
B.~Tuchming$^{3}$,      
D.-P.~van~der~Werf$^{17}$, 
D.~Won$^{12}$,          
S.~Wronka$^{2}$,        
P.~Yzombard$^{8}$       
}

\affiliation{$^{1}$ Institute for Particle Physics and Astrophysics, ETH Zurich, 8093 Zurich, Switzerland} 
\affiliation{$^{2}$ National Centre for Nuclear Research (NCBJ), ul. Andrzeja Soltana 7, 05-400 Otwock, Swierk, Poland} 
\affiliation{$^{3}$ IRFU, CEA, Université Paris-Saclay, F-91191 Gif-sur-Yvette, France} 
\affiliation{$^{4}$ The Pohang University of Science and Technology (POSTECH), Pohang, Republic of Korea} 
\affiliation{$^{5}$ Laboratoire Kastler Brossel, Sorbonne Université, CNRS, ENS-Université PSL, Collège de France, Campus Pierre et Marie Curie, 4, Place Jussieu, 75005, Paris, France} 
\affiliation{$^{6}$ Université d’Evry-Val d’Essonne, Université Paris-Saclay, Boulevard F.~Mitterand, 91000 Evry, France} 
\affiliation{$^{7}$ Université Paris-Saclay, CNRS/IN2P3, IJCLab, Orsay, France} 
\affiliation{$^{8}$ Conservatoire National des Arts et Métiers, 292 rue Saint Martin, 75003 Paris, France} 
\affiliation{$^{9}$ Université de Strasbourg, CNRS, IPCMS, UMR 7504, F-67000 Strasbourg, France} 
\affiliation{$^{10}$ Department of Physics, Stockholm University, Stockholm, Sweden} 
\affiliation{$^{11}$ Center for Underground Physics, Institute for Basic Science, Daejeon, Korea} 
\affiliation{$^{12}$ Department of Physics and Astronomy, Seoul National University, Seoul, Korea} 
\affiliation{$^{13}$ Department of Accelerator Science, Korea University Sejong Campus, Sejong, Korea} 
\affiliation{$^{14}$ Institute of Physics, University of Tokyo, Tokyo, Japan} 
\affiliation{$^{15}$ Institut Max von Laue - Paul Langevin (ILL), Grenoble, France} 
\affiliation{$^{16}$ CERN, EP Department, 1 Esplanade des Particules, 1217 Meyrin, Switzerland} 
\affiliation{$^{17}$ Department of Physics, Swansea University, Swansea, United Kingdom} 
\affiliation{$^{18}$ QUANTUM, Institut für Physik, Johannes Gutenberg Universität, Mainz, Germany} 
\affiliation{$^{19}$ {\emph Present Address:} National Metrology Institute of Japan (NMIJ), National Institute of Advanced Industrial Science and Technology (AIST), Tsukuba, Japan} 
\affiliation{$^{*}$ Corresponding author}
\emailAdd{regenfus@cern.ch}

\abstract{
The production of antihydrogen by the GBAR experiment at AD/ELENA requires good knowledge of the number of incident keV antiprotons, which can be problematic. We have used a commercial CMOS digital camera mounted around the experimental vacuum chamber to determine antiproton numbers from ionising particles created in the annihilation process on the surface of microchannel plate detectors which are used for beam imaging. We show that the multiplicity of emerging charged particles is as expected for individual annihilations of antiprotons with nucleons at rest, taking into account the surrounding material budget. Most of those particles are in the minimal ionising regime, but can be detected with nearly 100\% efficiency in the CMOS pixel detector, while due to the thin depletion layer the device is insensitive to background gammas. Thanks to the high granularity and small pixel size millions of antiproton annihilations can be reconstructed in a dense tracking environment over a large dynamic range with good resolution. From cluster length studies of non perpendicular tracks the thickness of the depletion zone and effective detection area was estimated. The cluster length also allows for a monitoring of track angles. Antiproton numbers are determined from the number of reconstructed clusters in the CMOS sensor by means of the covered solid angle relative to a calibration measurements with well known beam intensities at the most upstream location of the GBAR apparatus. Material effects on the emerging annihilation products were estimated by Monte Carlo (\geant) calculations, while annihilation artefacts on the complex surface of a microchannel plate are cancelled out in this approach. This method minimises largely systematic uncertainties, leading to a final error of roughly 10\%\ for the reconstruction of absolute antiproton numbers.}

\arxivnumber{xxxx.xxxx} 

\begin{document}
\maketitle
\flushbottom

\section{Introduction}
\label{sec:intro}

CERN's AD/ELENA facility \cite{Elena} routinely delivers intense 100\,keV antiproton pulses to several experiments, including GBAR (Gravitational Behavior of Antihydrogen at Rest)\,\cite{Adrich:2023tua}. Those experiments, aiming for high precision studies on antihydrogen, require careful monitoring of low energy (keV) antiprotons, as in the case for the recent measurement of the antihydrogen (\Hbar ) cross-section in the GBAR experiment\,\cite{GBARX26}. Particle imaging is done routinely by using microchannel plate (MCP) detectors, delivering both, fast electric, as well as 2D optical signals, when combined with a phosphor screen. They are sensitive to ionising radiation from particles impinging on their front surface, as well as by creation of electrons anywhere in the channels, e.g.~by the annihilation products of antiprotons. However, problems can arise in signal linearity due to high ionisation densities in case of large numbers of antiprotons annihilating on small fractions of their surface. 

Semiconductor micro-pattern detectors have an important share in the sector of modern particle-physics detectors, thanks to high signal-to-noise read out of ionisation charge by transistor based micro-electronics. After the primary developments of silicon-based charge-coupled devices, micro-strip and pixel detectors, CMOS sensors more recently emerged. Thanks to the availability of sub-micrometer lithography technology, it became  possible to integrate the sensor, read out electronics and detector control (e.g.~a global shutter) in small pixel sizes without creating large dead areas or hence reduced quantum efficiencies. Thanks to the very small capacitances, a high signal-to-noise ratio can be achieved for thin detectors, which is very useful for low-mass particle tracking.   

In this paper we present results from a CMOS sensor built into a commercial BAUMER VCXG-51M digital photo camera that has been used to detect particles resulting from annihiliations of antiprotons on the surface of MCPs inside the beam vacuum system. The camera was operated without a lens, covered by a light-tight foil and mounted outside the vacuum chamber, facing the \pbar\ annihilation points at typical distances of a few tens of centimeters. A large fraction of the detected particles are close to the minimal ionisation limit, but create well detectable signal clusters in the thin depletion layer of the CMOS pixels. The CMOS chip, designed to detect low-level visible light, is a 5 megapixel sensor, type Sony IMX264LLR\,\cite{Sony}, with $2448 \times 2048$ active square pixel cells in a 3.45\mum\ pitch, corresponding to a maximal active area of 59.7\mmsq . In the regime of several millions of annihilating antiprotons, this method leads to typically hundreds to thousands of minimal-ionising tracks perpendicularly traversing the CMOS sensor. Due to the low noise, the detection efficiency is close to 100\% , while the occupancy for the sensor is below 0.1\%\ due to the large number of pixels. The derived antiproton flux is prone to systematic errors due to uncertainties in the target nuclei (annihilation pion multiplicities), imperfect description of surrounding material, detector inefficiencies or backscattered antiprotons. In the following, we discuss the multiplicities of created charged particles (Sec.~2) together with a short description of the annihilation process and the particle transport which was determined by Monte Carlo simulations (Sec.~3), before going in the details of the signal detection (Sec.~4) and determination of the antiproton flux (Sec.~5) by means of a calibration measurement with known \pbar\ numbers at the location of the beam entrance into the GBAR experiment.

\section{Charged particle multiplicities in antiproton annihilations}
\label{sec:chrgdmult}

The annihilation of low energy antiprotons on material surfaces is understood by the subsequent steps of initial energy loss due to excitation and ionisation within the  atomic shells of the surface material, capture and cascading down to inner orbitals overlapping with the nucleus, and the final annihilation process. Antiproton annihilations on free protons produce on average close to three charged (\pipm ), and two neutral (\piz ) pions , the exact values being 3.05$\pm$0.04 and 1.93$\pm$0.12 \cite{Klempt}. Regarding annihilations on larger nuclei, in most cases antiprotons will annihilate with a proton or a neutron on the surface, leaving the rest of the nucleus as a spectator. Annihilations on neutrons produce, on average, more negative pions, leading to an overall ratio of roughly 7:5 for \pim /\pip\ abundances in nuclear annihilations.
\begin{figure}[htbp]
\centering
\includegraphics[width=0.9\textwidth]{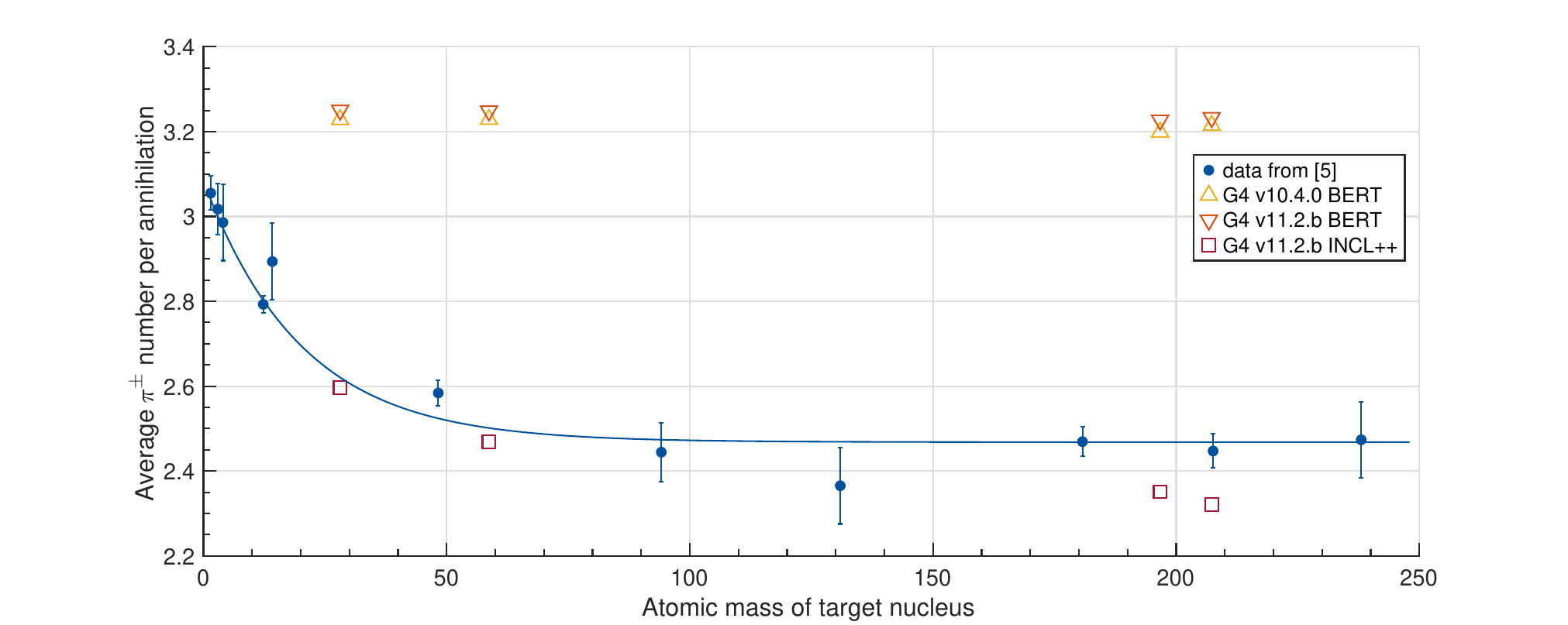}
\caption[]{Charged pion multiplicity versus target nucleus mass for \pbar\ annihilation data\,\cite{Bendiscioli} and several Monte Carlo results from this work. The line represents the best fit to an exponential decrease to a value of 2.468\,\cite{Bendiscioli}.}
\label{fig:Pi_mult_overv} 
\end{figure}
Created pions can be absorbed in the spectator nucleus with increasing probability of its size, leading to a reduction in the (average) rate of emerging charged pions from the value 3 to roughly 2.5 per annihilation for nuclei with A\,$\geq$\,60. This is illustrated from reference data collected in a 1994 review\,\cite{Bendiscioli} and displayed in Fig.\,\ref{fig:Pi_mult_overv} as solid circles with error bars. The line represents the best fit to an exponential decrease to a value of 2.468 (see\,\cite{Bendiscioli} for details). The open triangles are Monte Carlo simulation results from this work and are explained below. To derive precise numbers of antiprotons by counting annihilation tracks a good knowledge of the target material is hence required, which can be difficult. Furthermore, the number of emerging charged particles is affected by  surrounding material due to pair production by \piz\ gammas, pion absorption, particle decays, and other types of particle interactions. 

The description of antiproton annihilations on MCP surfaces is a complex task, due to the variety of used materials, tiny geometrical features (holes), as well as complex chemical modifications on the surface caused by reduction with hydrogen in the production process. The base material, a lead-silicate glass alloy, is often mixed with other oxides such as K$_2$O, Na$_2$O, Rb$_2$O, BaO, MgO or Bi$_2$O$_3$, in different ratios to facilitate the hydrogen reduction. The treatment improves the bulk resistance and greatly enhances secondary electron emission properties by creating a semiconductive surface layer. Depending on process temperature and application time, complex metallic molecular crystal structures are formed on the surface (e.g.~Pb$_7$Bi$_3$\,\cite{Zhang_2018}). MCPs can adsorb rest gas molecules (e.g.~water) on their surface, which can be difficult to be removed.  This can alter electric properties and yields uncertainties on the nature of the final annihilation target. 

A precise description of the \pbar\ annihilation process on a MCP surface is therefore difficult, since pions can be generated from annihilation on heavy or light elements that differ substantially in the generated charged particle multiplicity. Moreover, keV antiprotons can re-scatter off the surfaces of the contact metallisation in between the micro pores, with badly known mechanical properties. Here, we assume that antiprotons entering the holes (typ.~50-70\% surface ratio) will not scatter back and be trapped inside, eventually annihilating. Experimental and simulation efforts are ongoing in the collaboration to study this effect in more detail\,\cite{BSC26}.

The vacuum chamber material in the GBAR experiment, mainly consisting of stainless steel (SS) components of beam pipes and flanges, enlarge the number of apparent charged tracks measured outside the experimental apparatus due to pair production, which is a more dominant effect than the absorption of pions. On a 25\% error level, as a rough estimate under typical conditions in the experiment, an average value close to 5 is found for the multiplicity of charged particles emerging into 4$\pi$ per annihilating antiproton.

\section{Monte Carlo simulations}\!
\label{sec:MC}
\subsection{Model description}

The GBAR experiment aims to produce antihydrogen atoms (\Hbar ) and antihydrogen ions (\Hbarplus ) in charge-exchange reactions of keV antiprotons from ELENA with eV energy positronium atoms (a detailed description can be found e.g.~in\,\cite{Adrich:2023tua}). The experiment is roughly a 10\,m long assembly of beam optical components designed to transport an intense antiproton bunch ($\sim\!\!10^6$) with low divergence through a narrow reaction cavity (1.5 mm high by 1.0 mm wide by 20 mm long) where the positronium atoms are produced and confined. At several positions along the beamline, MCP detectors can be inserted to monitor \pbar\ beam profiles. In this work we discuss two positions in detail where the CMOS \pbar\ flux detector is employed: the upstream calibration position close to the beam entrance into the GBAR experiment (\MCPu), as well as the downstream location (\MCPd ) after the positronium target, where produced antihydrogen atoms are detected, or the number of available antiprotons has to be determined with highest precision.

\begin{figure}[htbp]
\centering
\includegraphics[height=54mm]{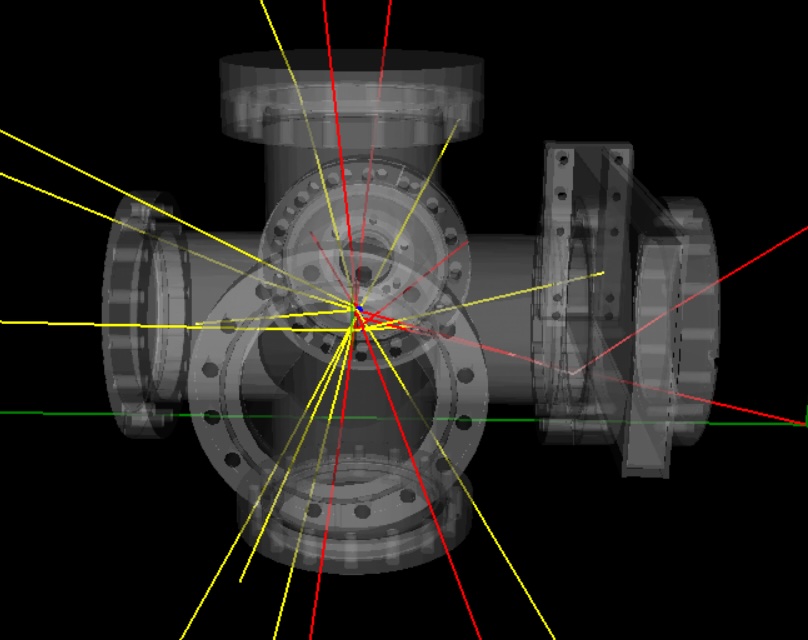}
\hspace{3mm}
\includegraphics[height=54mm]{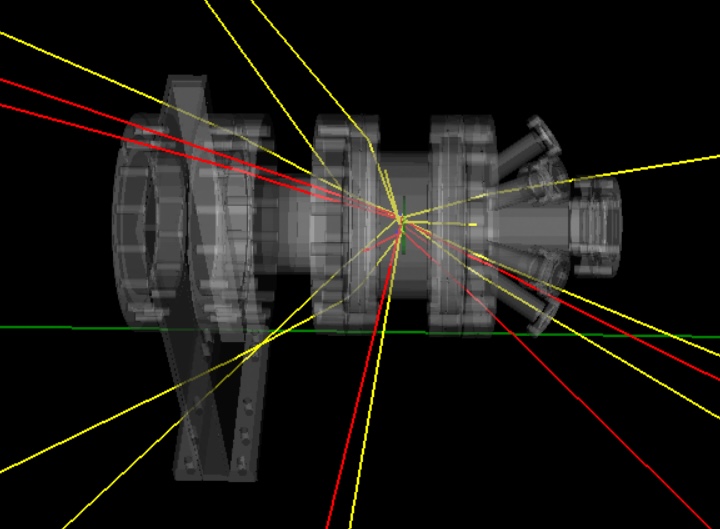}
\hspace{3mm}
\caption{MC simulation of annihilating antiprotons on \MCPu\ (left) and \MCPd\ (right). The translucent greyish components illustrate the precise description of material distributions in the simulation. The yellow and red tracks correspond to negative and positive charged pions, respectively.}
\label{Fig:MCsetups}
\end{figure}

The material distributions from both locations were taken from CAD data and fed into the \geant\ Monte Carlo (MC) framework\,\cite{Agostinelli:2002hh}. Figure\,\ref{Fig:MCsetups} left and right show event displays with an illustration of the surrounding beam pipe, including the tracks of a few annihilating antiprotons at the locations of \MCPu\ and \MCPd , respectively. For simplicity the MCP detectors were described as solid discs (of 1\mm\ thickness) with a composition of Si, Pb, Ni and Au elements in realistic ratios, the latter two form part of the thin surface contact metallisation. 

In general one million events were generated per Monte Carlo campaign. In a first approach the MC framework was tested for a description of \pbar\ annihilation multiplicities for the \geant\ versions 10.4.0 and 11.2.b with the BERT (Bertini-like cascade model) physics list which is known for being optimised for modelling \pbar\ annihilation processes at higher energies. 
In fact only a poor description for the pion multiplicities of \pbar\ annihilations on four selected target nuclei (Si, Ni, Au and Pb) is achieved, regardless of which \geant\ version is employed. The results are displayed as open triangles in Fig.\,\ref{fig:Pi_mult_overv} and were extracted from a small virtual sphere closely enclosing  the annihilation location to avoid material effects. The error for a description of \pbar\ annihilation multiplicities in this MC framework can reach 25\%.

Due to strong interest from the AD/ELENA experimental community for low-energy antibaryon interactions, recent efforts have led to an extension in the simulation code for low-energy antiproton annihilations based on the Li{\`e}ge IntraNuclear Cascades modeL (INCL) \cite{David:2025uxk}, 
which is available in the C++ version starting for \geant\ version 11.2.b by the INCL++ physics list. Results, obtained in this way for the same target nuclei as above, are displayed as open squares in Fig.\,\ref{fig:Pi_mult_overv} and demonstrate the substantial improvement in performance. Despite an ongoing problem with the non-isotropic emission of certain generated pions it was decided to use this code as a base for this work\,\footnote{Parts of these simulations were done in the framework of a Bachelor thesis at ETHZ\,\cite{Felix:2024}}. As described below this drawback does not affect this work, since only the ratio of MC simulations is used to derive the final result. 

\subsection{MC Results}
\label{sec:MCResults}

The average number of emerging charged particles per \pbar\ annihilation \Ntrppb\ was calculated in the MC simulation for the experimental configurations of \MCPu\ and \MCPd .
Typically, one million events were generated per run. Between 0 and 20\% of the antiprotons were found not to annihilate on the discs, scattering off a (solid) MCP surface.
The effect was found to be energy dependent and not clearly reproducible in the MC framework. Since we expect about 60\% of the antiprotons will enter the MCP holes and annihilate, we forced the annihilation to take place on the disc surface and attributed a systematic error to the scattering effect. As results are determined relative to the calibration measurement at \MCPu , only the ratio of this correction will enter calculations so the uncertainty is largely mitigated. Once charged particles are produced in the annihilation process, track multiplicities are derived from the number of hits \Ntr\ created in the CMOS sensor, depositing a detectable signal in a 10\mum\ depletion layer, weighted by the solid angle of the CMOS sensor seen from the location of the beam axis at the surface of the MCP. Further-on, possible detector inefficiencies or data reconstruction artefacts, like dead zones, or thresholds, cancel out in the ratio. The calculated numbers for each campaign of 6 runs (6M events) with statistical errors are shown in Table \ref{tab:MCresults}. The solid angles were calculated for the 59.7\mmsq\ detector area and distances of 247 and 340\mm\ for \MCPu\ and \MCPd , respectively. 
\begin{table}[h]
  \centering
    \begin{tabular}{|c|c|c|c|c|c|}
      \hline   
      Version & Setup & Solid angle & \Ntr & \Ntrppb & Ratio \MCPd\,/\,\MCPu \\
      \hline        
      G4 v11.2b & \MCPu & 77.9$\cdot 10^{-6}$ & 2494 & 5.338$\pm$0.107 & \multirow{2}{*}{1.046$\pm$0.035} \\
         INCL++ & \MCPd & 41.1$\cdot 10^{-6}$ & 1377 & 5.584$\pm$0.151 &\\
      \hline        
      G4 v10.4  & \MCPu & 77.9$\cdot 10^{-6}$ & 2738 & 5.860$\pm$0.112 & \multirow{2}{*}{1.008$\pm$0.033} \\
           BERT & \MCPd & 41.1$\cdot 10^{-6}$ & 1456 & 5.905$\pm$0.155 &\\
      \hline
    \end{tabular}%
  \caption{MC simulation results of each 6M \pbar\ annihilations on \MCPu\ and \MCPd\ employing the two \geant\ versions with two different physics lists (see text). The results from the older 10.4 G4 version with the BERT physics list are also displayed but not used for analysis. The uncertainties are purely statistical.}
  \label{tab:MCresults}
\end{table}

The experimentally determined value for the average multiplicity per \pbar\ annihilation \Ntrppb\ at the upstream calibration position of \MCPu\ (see section\,\ref{sec:calflux}) of 5.126\,$\pm$\,0.130 can be used for a first inspection of the simulation results. The data obtained by means of the G4 10.4 version overestimates this average track multiplicity by about 15\%. Data generated by the INCL++ class give a much more satisfactory result being only about 4\% larger than the \MCPu\ measurement. This difference can be attributed to an incorrect description of the target nuclei, inefficiencies in the CMOS detector or back-scattered antiprotons. 
As mentioned above, uncertainties are reduced in this data-driven approach using the experimentally determined flux at \MCPu\ for calibration and the ratio \MCPd /\MCPu\ (1.046) for the reconstruction of the \pbar\ flux at the position of \MCPd\ (see also section\,\ref{sec:calflux}). In this case the MC simulations are only needed to describe the differences in the particle transport at the two MCP positions.

\section{CMOS detector properties}
\label{sec:meas}

Primarily the CMOS detector is employed for a precise determination of the \pbar\ number passing the antihydrogen target region and impinging on \MCPd , where the downstream  MC model was created. For the evaluation of general sensor properties we used a more upstream position where the camera could be mounted at a larger distance ($\sim$60\cm ) to the beam axis to probe the device with better defined parallel tracks. 

\begin{figure}[ht]
\centering
\includegraphics[height=5cm]{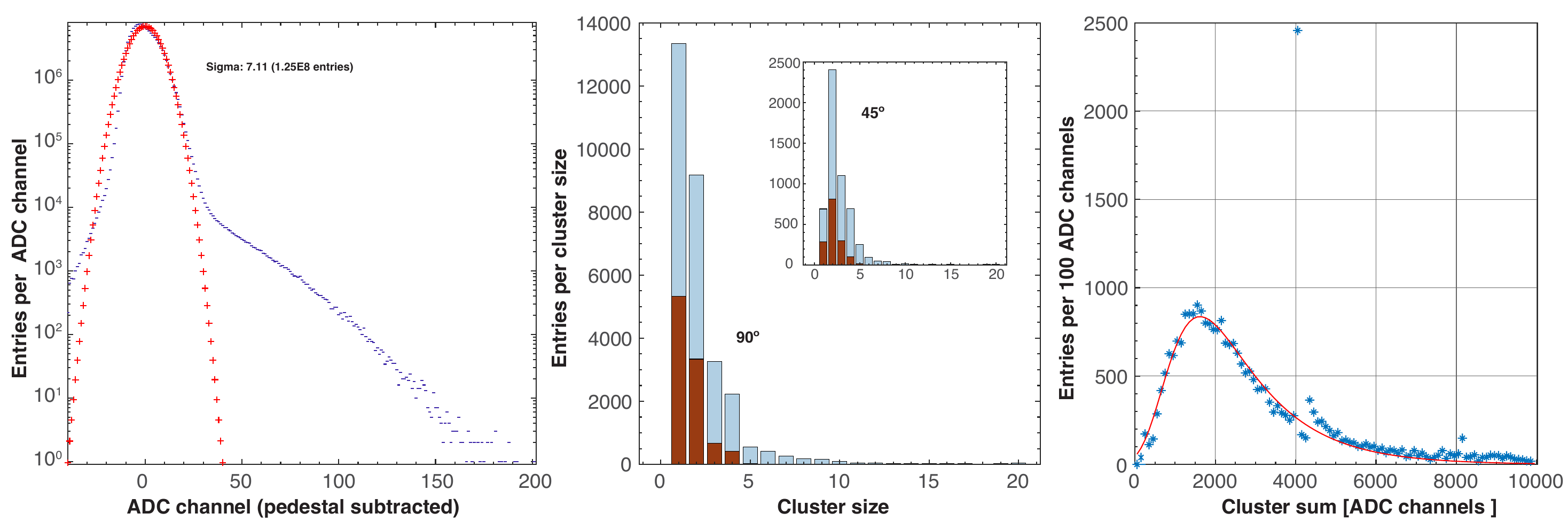}
\caption{Left:~Pixel signal distribution (dark events) and Gaussian fit ($\sig\!\approx\,$7\,ADC\,counts); Centre:~Cluster size histograms for signals from perpendicular tracks with (orange) and without (blue) cut on the  cluster sum; Right:~Cluster sum histogram of all reconstructed hits in the CMOS detector (the artefacts at 4096 and 8192 channels are due to overflow of the 12~Bit ADC) and the Landau-like fit (in red) to the data.}
\label{Fig:MultLand}
\end{figure}

The readout of the CMOS sensor in the BAUMER VCXG-51M camera is fully controlled by software from the main GBAR DAQ, based on the National Instruments LabView framework. All pixels are selected to be read out (no disable mask) by a global electronic shutter triggered on the arrival of antiprotons. Usual settings for exposure time and gain were 200\mus\ and 16, respectively. Other exposure values (e.g.~500\mus ) were also used differing slightly in pedestal values or numbers of noisy pixels. The readout is realised in an integrated 12 bit ADC architecture (4096 channels). From the CMOS sensor data sheet\,\cite{Sony} we find a single-pixel saturation capacity of about $10^{4}$ electron charges with a typical dark noise of 2.15\,\enrms . Unfortunately neither the thickness of the depletion layer, nor the fill factor is given. With those settings we extract a general rms noise of about 7\,ADC counts from pedestal data for more than 99\% of the pixels, while a small fraction ($\sim\!$44 channels) exhibit some higher noise up to 30\,ADC counts rms (see left panel of Fig.\,\ref{Fig:MultLand}). To find the number of detected tracks \Ntr\ in the CMOS sensor we take the raw data array of the 5M pixels and subtract the pedestal. All pixels above a common 5\sig\ threshold (150, derived from the more noisy pixels) are then flagged as fired and clusters are found from groups of adjacent pixels. This threshold creates on average less than 1 count per dark read-out and generates only a very small bias ($<$2\% of the area) on the pulse height distribution of reconstructed events. 
Figure\,\ref{Fig:MultLand} shows the histograms of the cluster sizes \nclu\ (number of pixels in a cluster) and cluster sums \Sclu\ (sum of pixel signals in a cluster) for about 25,000 reconstructed hits from 78 \pbar\ spills. The cluster sum histogram, corresponding to the total energy deposit in the depletion layer, is fitted by a Landau-like function (scaled and shifted Moyal) to determine the most probable (1650) and mean (2550) values of the distribution. The distribution is slightly distorted around 4096 and 8192 channels due to individually saturating pixels (12~Bit ADC limitation), but with no influence on the number of reconstructed clusters (this could be mitigated by reducing the gain).
For (mostly) perpendicular tracks the mean cluster size was found to be about 2.2. If only tracks with $\Sclu\!<\!3000$ are considered \nclu\ is reduced to 1.6 since events from the Landau tail with a larger dispersal of ionisation charge are removed. These values are well understood in terms of charge sharing among adjacent pixels\,\footnote{By geometric origin or diffusion, due to the small pixel cell size of 3.45\mum} and the effect of electron delta rays. 

In data taken with a tilted CMOS sensor we observe elongated clusters with increased values for \nclu\ and \Sclu , with no significant change in the number of reconstructed tracks per solid angle. Thus, in this work we estimate the detection efficiency of the CMOS sensor $>$98\% , being only determined by the threshold in the cluster reconstruction algorithm. From the cluster lengths of larger angle tracks we find a thickness of the depletion layer of roughly 10\mum . With the intrinsic properties of silicon this value corresponds to about 1000\,\en\ being produced per MIP. The mean of the Landau-fit scales nicely with the path length of the tracks in the sensor for the inclined data.  

\section{Calibration of the antiproton flux at the position of main MCP detector }
\label{sec:calflux}

\begin{figure}[ht]
\centering
  \centering
  \includegraphics[height=5.5cm]{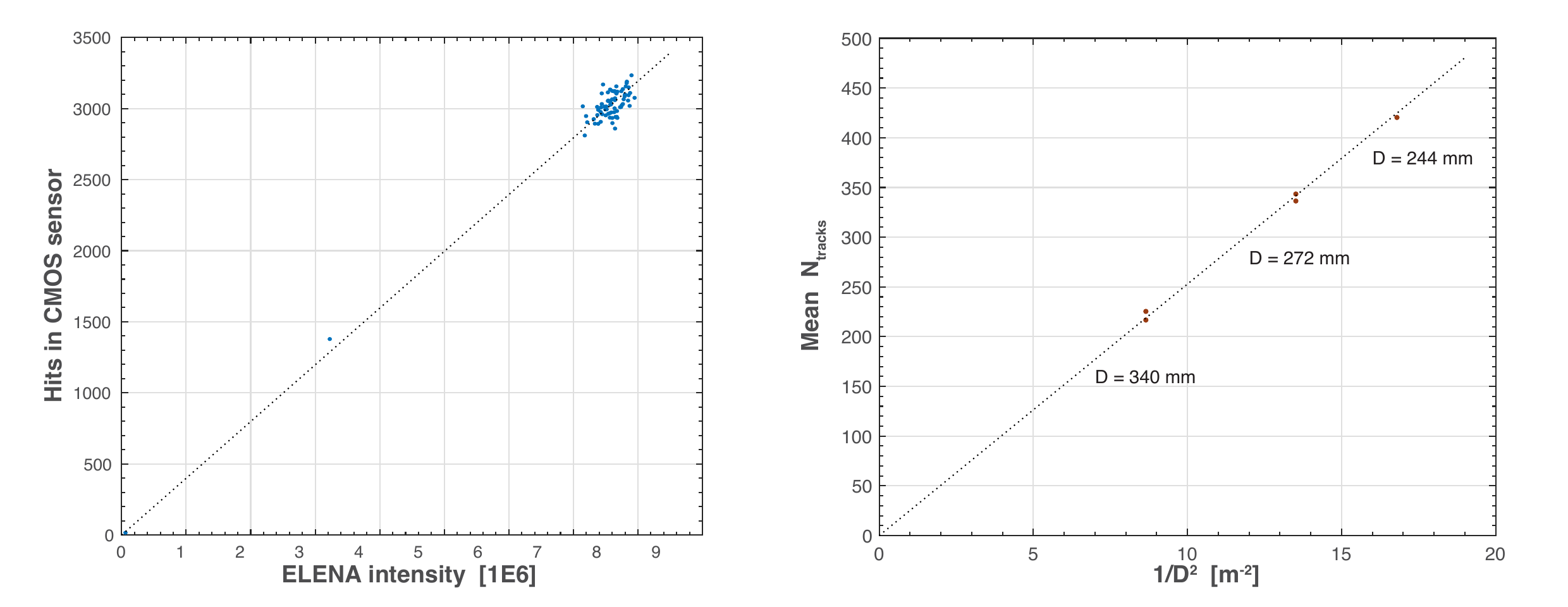}\hspace{8mm}
  \caption{Left: Number of hits reconstructed in the CMOS sensor at the position of \MCPu\ versus the antiproton flux derived from the ELENA induction monitor. The dashed line corresponds to a linear fit. Right: \Ntr\ in the CMOS sensor at \MCPd\ normalised to the ELENA intensity at various measurement distances D.}
  \label{fig:calibmeas}
\end{figure}

The main MCP detector after the positronium target in the experiment (\MCPd ) is used to detect created \Hbar\ atoms, as well as to determine the number of the antiprotons (\Npbar ) available for the creation of antihydrogen. The latter is calculated from the number of measured tracks \Ntr\ in the CMOS, when the \pbar\ beam is fully annihilating on the surface of \MCPd , using the apparent solid angle and the value for the effective number of tracks created per \pbar\ at this location in the experiment. With the area \Asens\ of the CMOS sensor and distance \Dist\ to the centre of the \pbar\ annihilation point on the surface of the MCP we find:  
\begin{equation*}
\Npbar = \Ntr\cdot 4\cdot\pi\cdot \Dist^{2} / (\Ntrppb\cdot\Asens) 
\label{Equ:flux}
\end{equation*}
This simple relation is valid thanks to large particle momenta and few material in the flight path. The 1/\Dist$^2$ dependence was verified at different detector distances (see Fig.\,\ref{fig:calibmeas}, right), where the data is averaged over each 20 shots for 5 campaigns and displayed versus 1/\Dist$^2$. Good general agreement is achieved with typical fluctuations on a 5\% level, presumably due to intensity variations of antiprotons extracted from the \pbar\ trap.

\Ntrppb$_{\MCPd }$ , the number of tracks created per \pbar\ annihilation at the position of \MCPd , is calculated from the experimental value \Ntrppb$_{\MCPu\,exp}$ at the position of \MCPu\ and the ratio of the MC simulations at those positions. As mentioned above the value for \Ntrppb$_{\MCPu\,exp}$ was determined from a comparison to an induction beam monitor in our beam line and from data recorded on \MCPu . For this calibration measurement the CMOS sensor was mounted at a distance of 24.7\,cm from the beam axis behind a CF150 blind flange. Figure\,\ref{fig:calibmeas} to the left shows the scatter plot of reconstructed hits in the CMOS versus the number of antiprotons from several ELENA spills with a linear fit to the data (dashed line), resulting to the value
\begin{equation*}
\Ntrppb_{\MCPu\,exp}\,=\,5.126\,\pm\,0.130
\label{Equ:cal}
\end{equation*}
the average number of reconstructed tracks in the CMOS detector per impinging \pbar . This value is used for the calibration. The precision of the induction beam monitor was verified to be better 5\% by measuring the total electric charge of \Hminus\ bunches from ELENA fully absorbed on the surface of \MCPu\ (under a positive bias voltage to avoid secondary electron emission). 
We also compared to a second induction beam monitor at a different location in the ELENA beam line and found less than 0.4\% deviation between the two ELENA sensors. Finally for the value for \Ntrppb$_{\MCPd }$ we achieve 
\begin{equation*}
\Ntrppb_{\MCPd }  = \Ntrppb_{\MCPu\,exp}\,\cdot\, 
{\rm MC\,ratio\,_{\MCPd /\MCPu }}\,=\,5.126\cdot1.046\,=\,5.362\,\pm\,0.225
\label{Equ:Ntr5corr}
\end{equation*}
with a combined error of 4.2\% from the result of the linear fit and the ratio of the MC results from table\,\ref{tab:MCresults} (statistical). Using the standard measurement distance \Dist\,=\,340\mm\ we achieve a handy formula for a determination of antiproton numbers annihilating on \MCPd\ in the experiment:
\begin{equation*}
\Npbar = \Ntr \cdot [4538 \pm 195(stat) \pm 380(sys)]
\label{Equ:handyflux}
\end{equation*}
The systematic error of 8.4\% is estimated from the incoherent sum of different contributions:  Differences derived from the MC simulations in the re-scattering between \MCPu\ and \MCPd\ (5\%)\,\footnote{A recent study in the same project confirmed the validity of this estimate\,\cite{BSC26} }, the ELENA induction beam sensor (4\%), errors in the placement distance (3\%) and beam fluctuations (5\%) caused by instabilities in the experimental environment. Inefficiencies in the CMOS detector or similar problems cancel out in this approach.

\section{Conclusions}
\label{sec:concl}

We employ a CMOS sensor from a commercial digital image camera as a detector for antiproton annihilation products to reconstruct antiproton numbers. The annihilations take place inside the experimental vacuum apparatus of the GBAR experiment on the surface of beam imaging MCPs. Thanks to the very low readout noise and large number of pixels the measurement is nearly 100\% efficient and linear over a large range. Millions of annihilating antiprotons can be reconstructed with ease. Systematic errors stemming from pion annihilation multiplicities, detector inefficiencies or other measurement problems are mitigated in calibrating the CMOS detector to a known \pbar\ beam flux. In this way a reconstruction of \pbar\ numbers with a total error of about 10\% is achieved.      

\section*{Acknowledgements}
\label{sec:ackns}

We thank L. Ponce and the AD/ELENA team as well as F. Butin and the CERN EN team for
their fruitful collaboration. We also thank A. Beynel for his continued support with the alignment
of the \pbar\ trap. This work is supported by: JSPS KAKENHI Grant-in-Aid for Scientific Research A
20H00150 and Fostering Joint International Research A 20KK0305 (Japan), 
the Action Thématique GRAM of CNRS/INSU with INP and IN2P3 co-funded by CNES (France),
SPHINX ANR-22-CE31-0019 (France), ESPRIT ANR-22-CE30-0028-01 (France), BESCOOL ANR/DFG ANR-13-ISO4-0002-01 (France/Germany), the Swiss National Science Foundation (Switzerland) grants
197346, 216673 and 232699 and ETH Zurich (Switzerland) grant ETH-46 17-1, the Swedish
Research Council (VR) grants 2017-03822, 2021-04005 and 2025-04378, the German cluster of excellence PRISMA, and the following grants from Korea:
IBSR016-Y1, IBS-R016-D1, UBSI Research Fund (No. 1.220116.01) of UNIST, POSTECH Initial
Settlement Support Fund, NRF-2016R1A5A1013277, NRF-RS-2022-00143178,
NRF-2021R1A2C3010989, and NRF-2016R1A6A3A11932936. The GBAR collaboration is an
International Research Network, supported by CNRS, France.





\bibliographystyle{JHEP}
\bibliography{biblio.bib}


\end{document}